\documentclass[review]{elsarticle}

\usepackage{hyperref}

\begin{document}

\begin{frontmatter}

\title{Response of microchannel plates in ionization mode to single particles and electromagnetic showers}

\author[Budker,NSU,NSTU]{A. Yu. Barnyakov}
\author[Budker,NSU]{M. Yu. Barnyakov}
\author[MiB]{L. Brianza}
\author[INFNRm]{F. Cavallari}
\author[INFNRm,Roma]{M. Cipriani}
\author[MiB]{V. Ciriolo}
\author[INFNRm,Roma]{D. del Re}
\author[INFNRm,Roma]{S. Gelli}
\author[MiB]{A. Ghezzi}
\author[MiB]{C. Gotti}
\author[MiB]{P. Govoni}
\author[Budker,NSU]{A. A. Katcin}
\author[MiB]{M. Malberti}
\author[MiB]{A. Martelli\fnref{nowatcern}}
\author[INFNRm,Roma]{B. Marzocchi}
\author[INFNRm]{P. Meridiani}
\author[INFNRm,Roma]{G. Organtini}
\author[INFNRm,Roma]{R. Paramatti}
\author[MiB]{S. Pigazzini}
\author[INFNRm,Roma]{F. Preiato}
\author[Budker,NSU]{V. G. Prisekin}
\author[INFNRm,Roma]{S. Rahatlou}
\author[INFNRm]{C. Rovelli\corref{mycorrespondingauthor}}
\author[INFNRm,Roma]{F. Santanastasio}
\author[MiB]{T. Tabarelli de Fatis}

\fntext[nowatcern]{Now at CERN}
\cortext[mycorrespondingauthor]{Corresponding author: chiara.rovelli@roma1.infn.it}

\address[Budker]{Budker Institute of Nuclear Physics, Lavrentieva 11, Novosibirsk 630090, Russia}
\address[NSU]{Novosibirsk State University, Pirogova 2, Novosibirsk 630090, Russia}
\address[NSTU]{Novosibirsk State Technical University, Karla Marksa 20, Novosibirsk 630073, Russia}
\address[MiB]{Universit\`a di Milano-Bicocca and INFN, Sezione di Milano-Bicocca, Piazza della Scienza 3, 20126, Milano, Italy}
\address[INFNRm]{INFN, Sezione di Roma, Piazzale A. Moro 2, 00185, Roma, Italy} 
\address[Roma]{Sapienza, Universit\`a di Roma, Piazzale A. Moro 2, 00185, Roma, Italy} 

\begin{abstract}
Hundreds of concurrent collisions per bunch crossing are expected at future hadron colliders.
Precision timing calorimetry has been advocated as a way to mitigate the pileup effects and,
thanks to their excellent time resolution, microchannel plates (MCPs) are good candidate
detectors for this goal. 
We report on the response of MCPs, used as secondary emission detectors, 
to single relativistic particles and to electromagnetic showers.
Several prototypes, with different geometries and characteristics, were exposed to particle beams at the 
INFN-LNF Beam Test Facility and at CERN. Their time resolution and efficiency 
are measured for single particles and as a function of the multiplicity of particles.
Efficiencies between 50\% and 90\% to single relativistic particles are reached, 
and up to 100\% in presence of a large number of particles. Time resolutions 
between 20~ps and 30~ps are obtained.
\end{abstract}

\begin{keyword}
\texttt{Microchannel plates, secondary emission, electromagnetic showers, time response, efficiency, calorimetry}
\PACS{29.40.Vj, 85.60.Ha, 79.20.Hx}
\end{keyword}

\end{frontmatter}


\section{Introduction}
\label{intro}

The projected beam intensity of future hadron colliders~\cite{HLLHC}\cite{FCC} will result in several hundreds of
concurrent collisions per bunch crossing, spread over a luminous region of a few centimeters along the beam axis and
of about a few 100~ps in time. The scientific program at these colliders, which includes precision characterization
of the Higgs boson, measurements of vector boson scattering, and searches for new heavy or exotic particles, will
benefit greatly from the enormous dataset. However, particle reconstruction and correct assignment to primary
interaction vertices at high vertex densities presents a formidable challenge to the detectors that must be overcome
in order to harvest that benefit. Time tagging of minimum ionizing particles (MIPs) and of neutral particles in the
calorimeters with a resolution of a few 10~ps provides further discrimination of the interaction vertices in the same
bunch crossing, beyond spatial tracking algorithms~\cite{Alice}.

Among other sensors that are being investigated for precision timing of charged tracks, microchannel plates
(MCPs)~\cite{mcp}, renowned for their fast response, have been advocated as a candidate detector to time tag electromagnetic
showers and MIPs and tested with moderate success~\cite{Derevshchikov}\cite{Bondila}\cite{Adams}\cite{Adams2}.
In a previous work~\cite{btf14}, we reported time resolutions of about 50 ps with cosmic muons and efficiencies of about 50\% in the 
detection of single relativistic charged particles. The detectors consisted
in a stack of two MCP layers with the dual function of seeding the cascade process, via the secondary electrons
extracted from the MCP surface by the incoming ionizing particle, and of providing signal amplification. Similar
studies~\cite{Apresyan1}\cite{Apresyan2}\cite{Apresyan3} reported results comparable to ours for MIPs and virtually 
full efficiency in the detection of electromagnetic showers, at shower depths where the track multiplicity is high, 
with time resolutions at the level of a few tens of ps.
In this paper, we further characterize the response of MCPs in the direct detection of ionizing particles, 
hereafter referred to as `ionization-MCPs' or shortly `i-MCPs'. 
The potential advantage of i-MCPs consists in the elimination of the photocathode, resulting in a easier and 
more robust construction and in a potentially larger radiation tolerance.
We report on the dependence of the i-MCP performance on the stack geometry (number of layers or aspect ratio)
and the use of MCP with high emissivity layers is also investigated. Several different prototypes of i-MCP detectors
were exposed to charged particles beams. 
After the description of the detectors and of the measurement setup~(Sec.~\ref{detectors} and Sec.~\ref{measurements}) 
we present results in terms of efficiency and time resolution in response to single particles~(Sec.~\ref{singleele}). 
We report also about the behaviour of i-MCP prototypes in response to electromagnetic showers at different depths~(Sec.~\ref{showers}). 
Due to their excellent time response, indeed, a layer of MCPs embedded in a calorimeter or in a preshower could be exploited to provide a precise time response for photons.

\section{Detectors description and operation}
\label{detectors}

The usage of MCPs as secondary electron emitters is investigated using either
sealed MCP devices 
developed at BINP (Novosibirsk) in collaboration with the Ekran FEP manufacturer~\cite{Barnyakov}
or layers by the Photonis and Incom manufacturers mounted inside a vacuum chamber
(with pressure kept below 10$^{-5}$~mbar using a turbo molecular pump).
The detectors are operated in `i-MCP mode', therefore when a photocathode is present a retarding bias 
is applied to the gap
between the photocathode itself and the MCP, as opposed to the standard `PMT-MCP' mode.
This prevents photoelectrons emitted from the photocathode from reaching the MCP surface and triggering an avalanche.
In this configuration, the response of the detector is uniquely determined by the secondary emission of electrons from the MCP
layers which are crossed by ionizing particles.

\begin{figure}[!htbp]
\begin{center}
\includegraphics[width=0.49\textwidth]{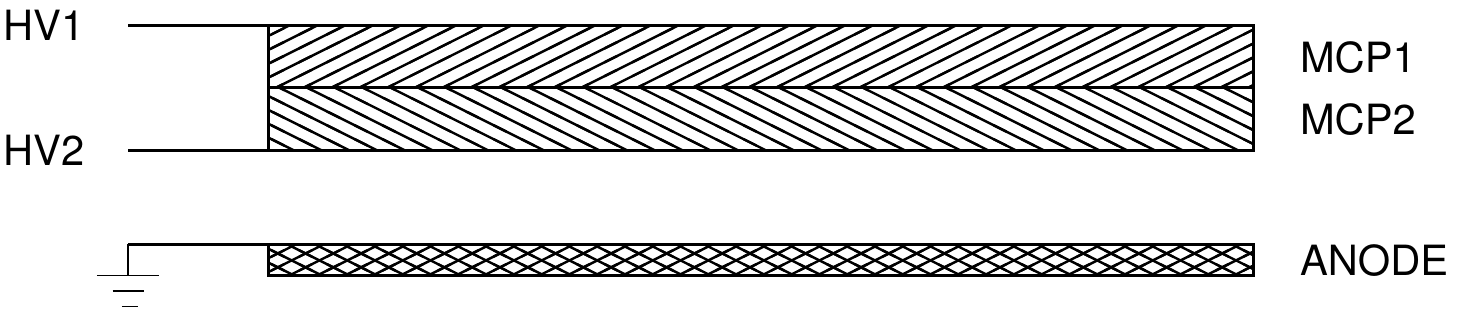}
\includegraphics[width=0.49\textwidth]{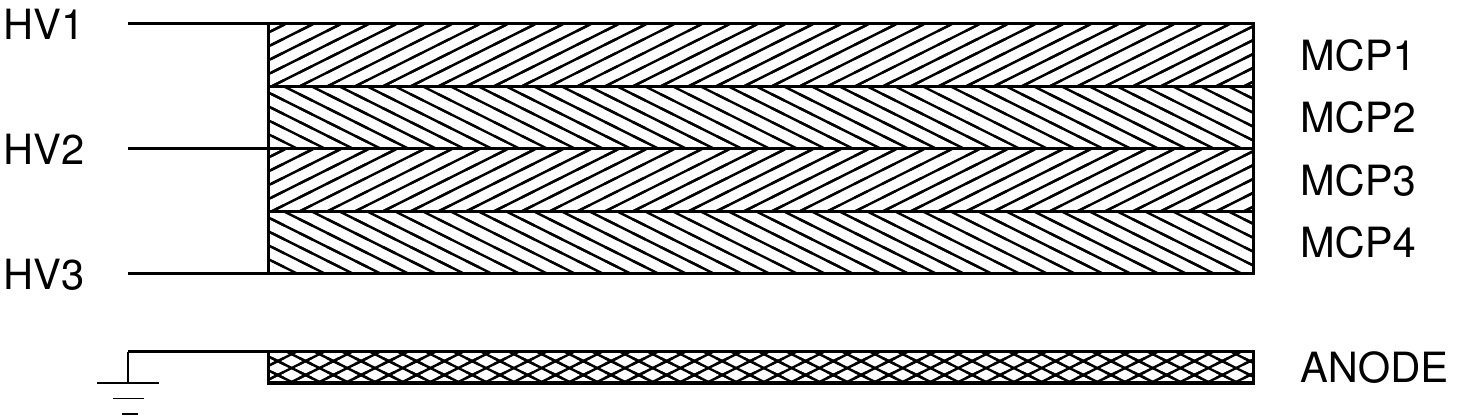}
\caption{Sketch of the two electrical configurations under test: without separation of the emission and amplification stages (left) and with separated stages (right).}
\label{fig:schemi}
\end{center}
\end{figure}

Devices are built with two, three or four layers of lead glass MCPs,
except for one prototype made of borosilicate glass coated with
emissive and resistive layers.
The channel diameter varies in the range from 3.5~$\mu$m to 25~$\mu$m, and its ratio to the
layer thickness (aspect ratio) ranges from 1:40 to 1:90 depending on the prototype.
The channels have a bias angle to the photodetector axis of a few degrees.
Two different electrical configurations are tested, which are sketched in Fig.~\ref{fig:schemi}.

In the first one, same as in~\cite{btf14}, a voltage divider is used to provide about 90\% of the
voltage drop through the entire MCP stack (2 or 3 layers). The layers carry out the dual function of seeding
the cascade process, via secondary electrons extracted from the MCP, and providing signal amplification.
In the second configuration the MCP layers are separated by the presence of electrodes,
dividing a stage where the emission happens with large probability from the amplification stage. 
More details on this approach will be given in the following.

The specifications of the detectors are detailed in Tab.~\ref{tab:detH4} and Tab.~\ref{tab:detBTF}
for the two configurations.
The prototypes which are mounted inside a vacuum chamber (VC)
have slightly less than one radiation length of material in front due to the steel flange closing the chamber.
\begin{table}[!h]
\begin{center}
\begin{tabular}[h]{|c|c|c|c|c|}
\hline
Device & Assembly & Layers & Diameter & Aspect ratio \\
\hline
\emph{40$\times$2}  & Sealed   & 2 & 12$~\mu$m  & 1:40 \\
\emph{40$\times$3}  & Sealed   & 3 & 8$~\mu$m   & 1:40 \\
\emph{SEE}          & Sealed   & 2 & 7.5$~\mu$m & 1:40 \\
\emph{MGO}          & VC       & 2 & 10$~\mu$m  & 1:40 \\
\hline
\end{tabular}
\caption{Devices without separation of the emission and amplification stages.
The designation given to each device indicates either the aspect ratio times the number of layers
or the superficial treatment applied (secondary emission enhancement \emph{SEE} or atomic layer deposition with 
Magnesium oxide \emph{MGO}).}
\label{tab:detH4}
\end{center}
\end{table}
\begin{table}[!h]
\begin{center}
\begin{tabular}[h]{|c|c|c|c|c|c|c|}
\hline
Device & Assembly & Layers & Em.Layer 1 & Em.Layer 2 & Ampl.Layer 1 & Ampl.Layer 2 \\
\hline
\emph{40$\times$2+40$\times$2} & Sealed & 4 & 7~$\mu$m, 1:40   & 7~$\mu$m, 1:40   & 7~$\mu$m, 1:40  & 7~$\mu$m, 1:40 \\
\emph{90$\times$1+40$\times$2} & Sealed & 3 & 3.5~$\mu$m, 1:90 & -                & 7~$\mu$m, 1:40  & 7~$\mu$m, 1:40 \\
\emph{90$\times$2+40$\times$2} & Sealed & 4 & 3.5~$\mu$m, 1:90 & 3.5~$\mu$m, 1:90 & 7~$\mu$m, 1:40  & 7~$\mu$m, 1:40 \\
\emph{80$\times$2+40$\times$1} & VC     & 3 & 25~$\mu$m, 1:80  & 25~$\mu$m, 1:80  & 25~$\mu$m, 1:40 & -      \\
\emph{80$\times$1+40$\times$1} & VC     & 2 & 8~$\mu$m, 1:80   & -                & 10~$\mu$m, 1:40 & -       \\
\hline
\end{tabular}
\caption{Devices operated with separated emission and amplification stages.
For each layer the channel diameter and the aspect ratio are quoted.
The designation given to each device indicates the aspect ratio times the number of layers for the emission stage (first part) 
and for the amplification stage (second part).}
\label{tab:detBTF}
\end{center}
\end{table}

\section{Experimental setup}
\label{measurements}

The i-MCP detectors were exposed to particle beams, in order to characterize their response
to single particles and electromagnetic showers in terms of detection efficiency and time resolution.

The devices without separation of the emission and amplification stages were tested at the H4~\cite{h4} and 
H2~\cite{h2} beam lines at the CERN North Area.
The electron beam at the H4 Area is extracted from the CERN SPS and can be tuned in the momentum range
from 10~GeV to 200~GeV. Our efficiency and time resolution measurements were performed with
20~GeV and 50~GeV electrons.
The MCP devices were mounted in a box with the optical window orthogonal to the beam direction.
Signals were read from a 32-channels digitizer (CAEN V1742) with 5~GHz sampling frequency.
A hodoscope was placed along the beam line upstream of the MCPs and readout into a gated-ADC,
for beam centering and offline selection purposes.
It consisted of four planes with 64 parallel scintillating fibers each, 1 mm in diameter and staggered,
covering an acceptance of 20$\times$20~mm$^2$ in the coordinates transverse to the beam.
One of the MCPs along the beam line was operated in PMT-MCP mode, to provide a trigger and a reference
for the efficiency and time measurements. 
The H2 beam is a secondary particle beam extracted from the CERN SPS providing hadrons,
electrons or muons of energies between 10~GeV and 360~GeV. In this case, our efficiency measurements were performed 
with 150~GeV muons with a similar setup as in H4.

\begin{figure}[!htbp]
\begin{center}
\includegraphics[width=0.60\textwidth]{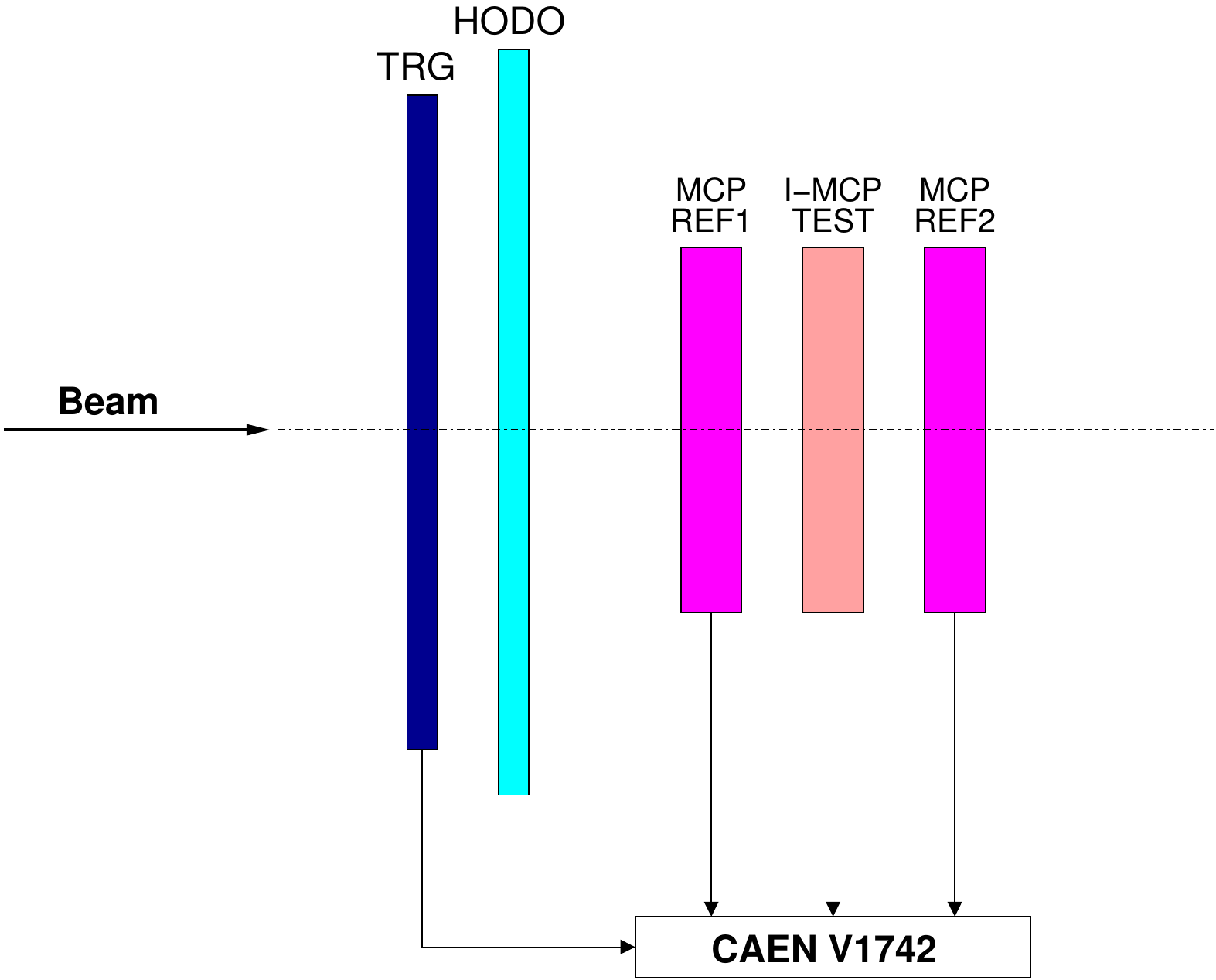}
\caption{Schematic view of the MCPs and other equipments position along the beam line at the BTF in Frascati.
A plastic scintillator counter (TRG) and a hodoscope (HODO) are put in front of the MCPs along the beam line.
The i-MCP under study (i-MCP TEST) is put between two MCPs operated in PMT-MCP mode 
used as references (MCP REF1, MCP REF2).
}
\label{fig:beamSetup}
\end{center}
\end{figure}

The devices with separated emission and amplification stages were tested at the T9 area at CERN~\cite{t9} 
and at the Beam Test Facility (BTF) of the INFN Laboratori Nazionali di Frascati (Italy)~\cite{btf}.
In the T9 Area a secondary beam with momentum range from 1~GeV to 5~GeV
is originated from the CERN PS beam impacting on a target. 
Our measurements were performed on a beam of 2~GeV electrons (20\%) and pions (80\%). 
The experimental setup was similar to the one used at the North Area and a 
Cherenkov detector was used to separate the electron and pion components of the beam.
The BTF in Frascati provides 10~ns long electron pulses with tunable energy (up to about 500~MeV), repetition rate (up to 49~Hz)
and intensity (from 1 to 10$^{10}$ particles per pulse). Our measurements were performed with 491~MeV electrons and an intensity
tuned to provide an average of about one electron per pulse. A similar beam line equipment as in H4 was used,
with only minimal differences. One MCP operated in PMT-MCP mode was placed upstream of the i-MCPs along the
beam line
and another one downstream, to select events going through the i-MCPs without showering.
Also, a 5~mm thick plastic scintillator counter with an area of 24$\times$24~mm$^2$
was installed in front of the MCPs, together with a hodoscope, to allow selecting events with single
electrons impinging on the detectors.
A schematic view of the BTF setup is shown in Fig.~\ref{fig:beamSetup}.

\begin{figure}[!htbp]
\begin{center}
\includegraphics[width=0.60\textwidth]{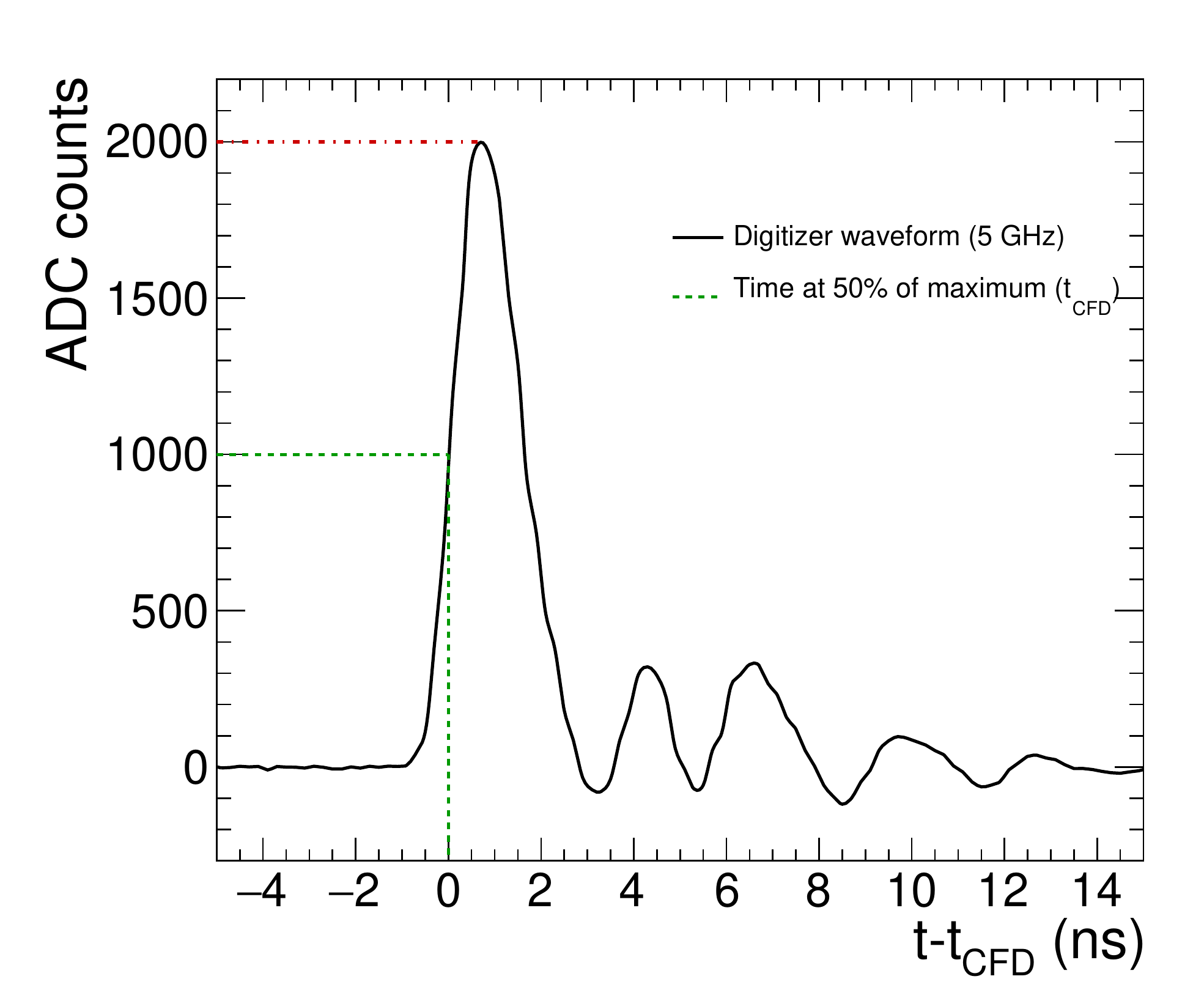}
\caption{A typical MCP waveform as extracted from the digitizer (solid line). 
On the x-axis the time $t_{\texttt{CFD}}$ for which the signal amplitude is 50\% (dashed line) of its maximum 
value (dashed-dotted line) is subtracted as an offset.}
\label{fig:pulseshape}
\end{center}
\end{figure}

The MCP devices produce very fast signals, with a rise-time of the order of 1~ns.
A typical MCP waveform is shown in Fig.~\ref{fig:pulseshape}, where the secondary peaks are due to
an imperfect matching between the anode and the transmission line to the digitizer.
For each event, a coincidence window of 60~ns is opened around the time corresponding to the maximum amplitude of the
MCP operated in PMT-MCP mode acting as a trigger. The maximum of the waveform of the i-MCP under test is 
searched inside this window. 
The signal time information is extracted from the interpolated waveforms via a constant fraction discriminator
(CFD), corresponding to the time when the amplitude is half of its maximum.

\section{Response to single charged particles}
\label{singleele}

To characterize the i-MCP detectors behaviour their efficiency and time resolution
were first measured in response to single electrons and muons.
Events consistent with a single particle are selected in the offline analysis
requiring a pulse larger than 200 ADC counts in the reference PMT-MCP device.
The hodoscope information is also used when available.
At the BTF, the bunched beam structure requires a further selection for single electron events,
identified from the pulse-height of the signal in the scintillator counter which is asked to be
between 200 and 700 ADC counts.
Furthermore, a signal between 200 and 1200 ADC counts is required in the PMT-MCP downstream of the
i-MCP (the upper limit is set to discard events in which the electron makes a shower along
the beam line).
\begin{figure}[!htbp]
\begin{center}
\includegraphics[width=0.45\textwidth]{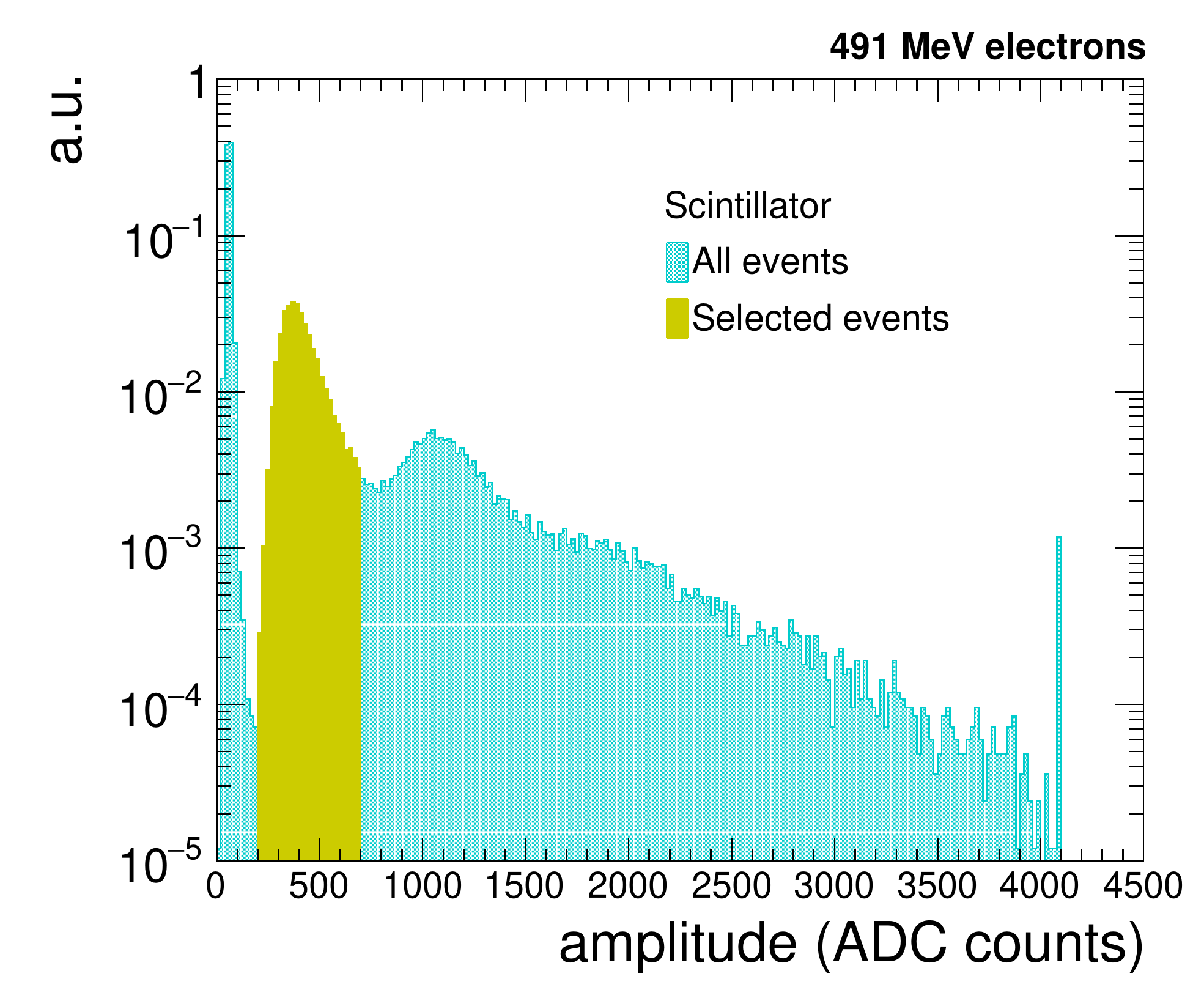}
\includegraphics[width=0.45\textwidth]{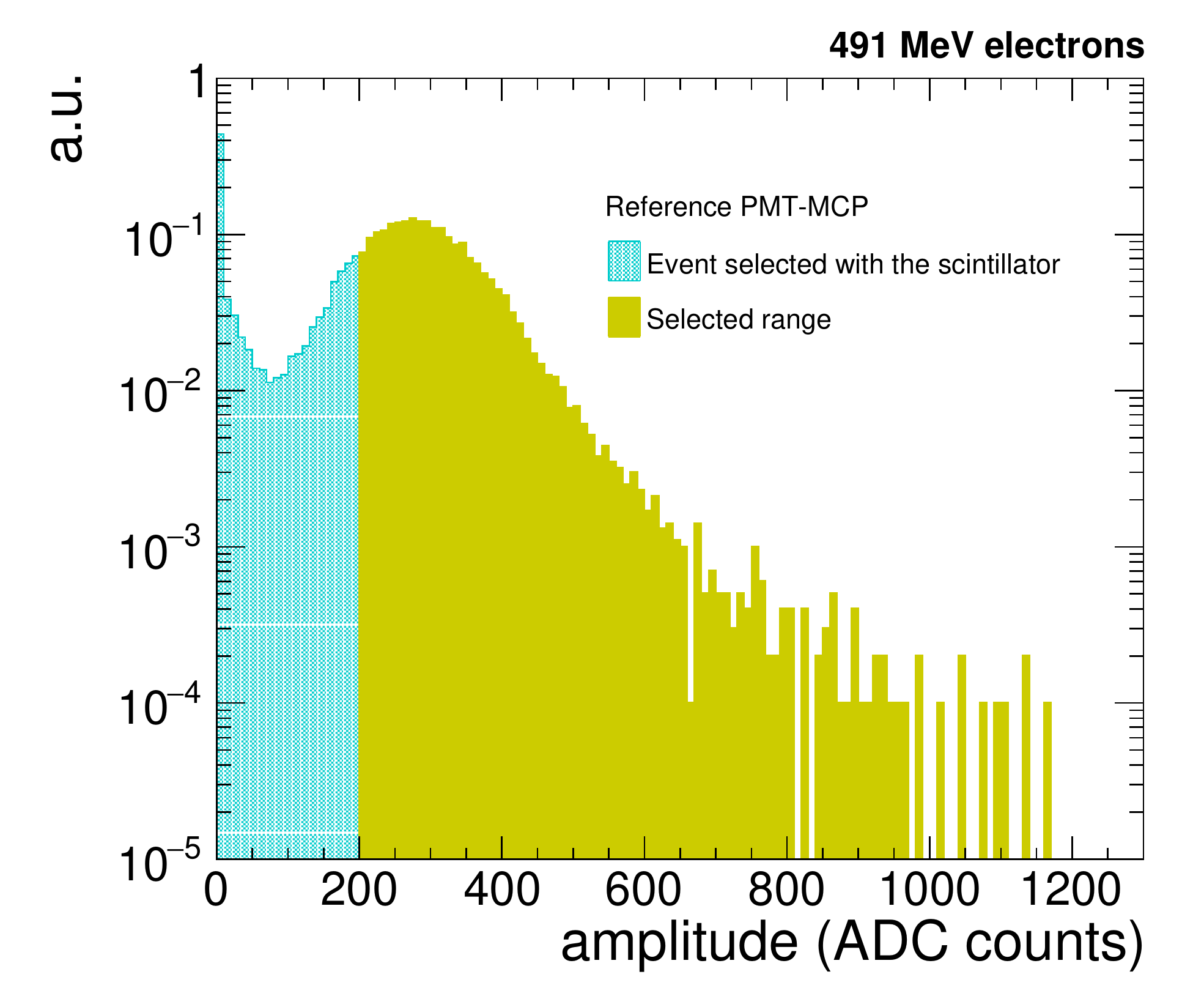}
\caption{Typical ADC spectrum of the scintillator counter (left) and of the MCP operated in PMT-MCP mode and put as a first detector along the beam line (right) at the BTF. In the right plot, only the events selected based on the scintillator response are shown. The region corresponding to single electron events is shown as a filled histogram in both plots. }
\label{fig:mib2}
\end{center}
\end{figure}
Typical spectra of the scintillator counter and of the PMT-MCP upstream of the 
test i-MCP at the BTF are shown in Fig.~\ref{fig:mib2}.

\subsection{Efficiency of single-stage MCP stacks}
\label{singlestage}

\begin{figure}[!htbp]
\begin{center}
\includegraphics[width=0.45\textwidth]{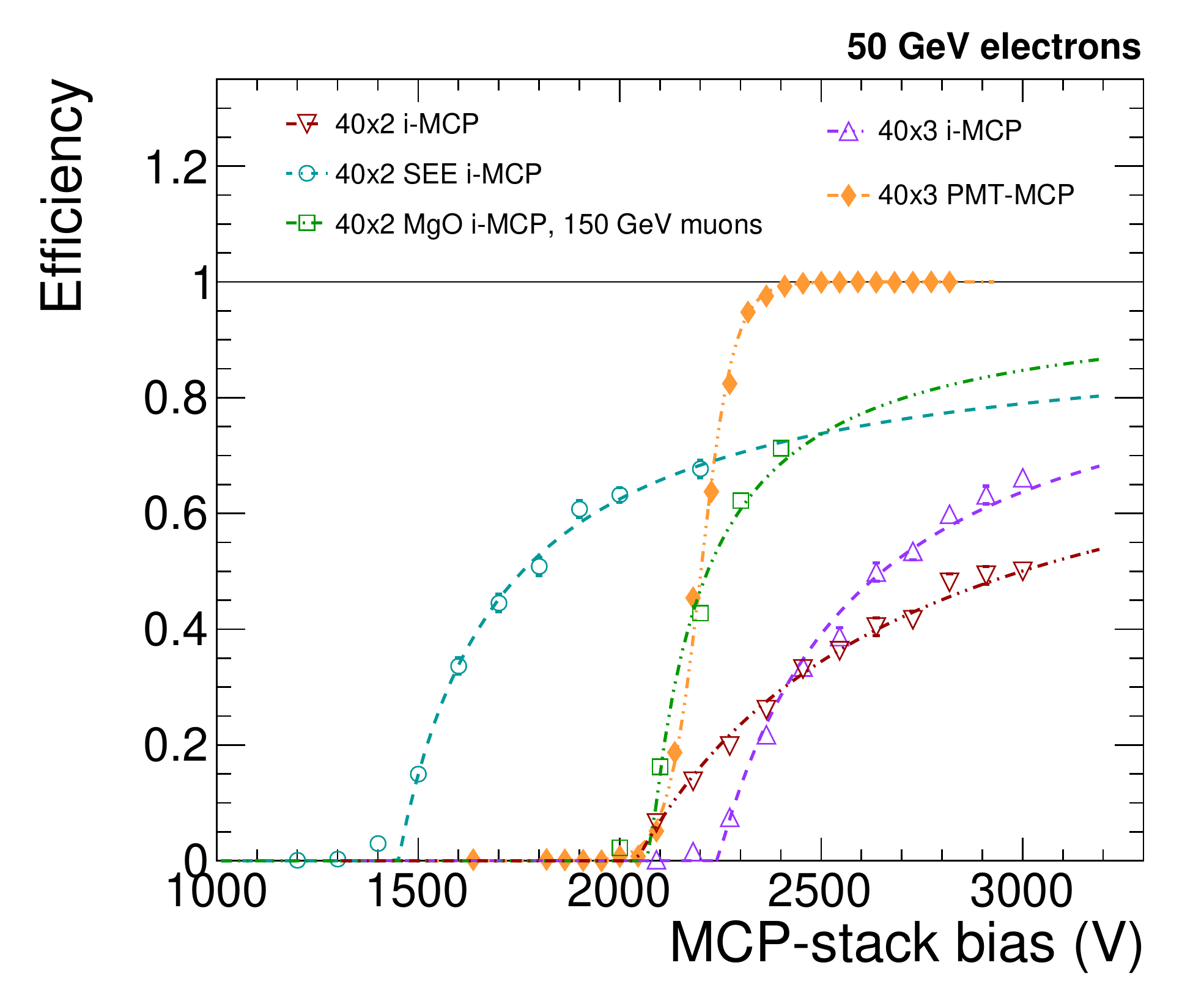}
\includegraphics[width=0.45\textwidth]{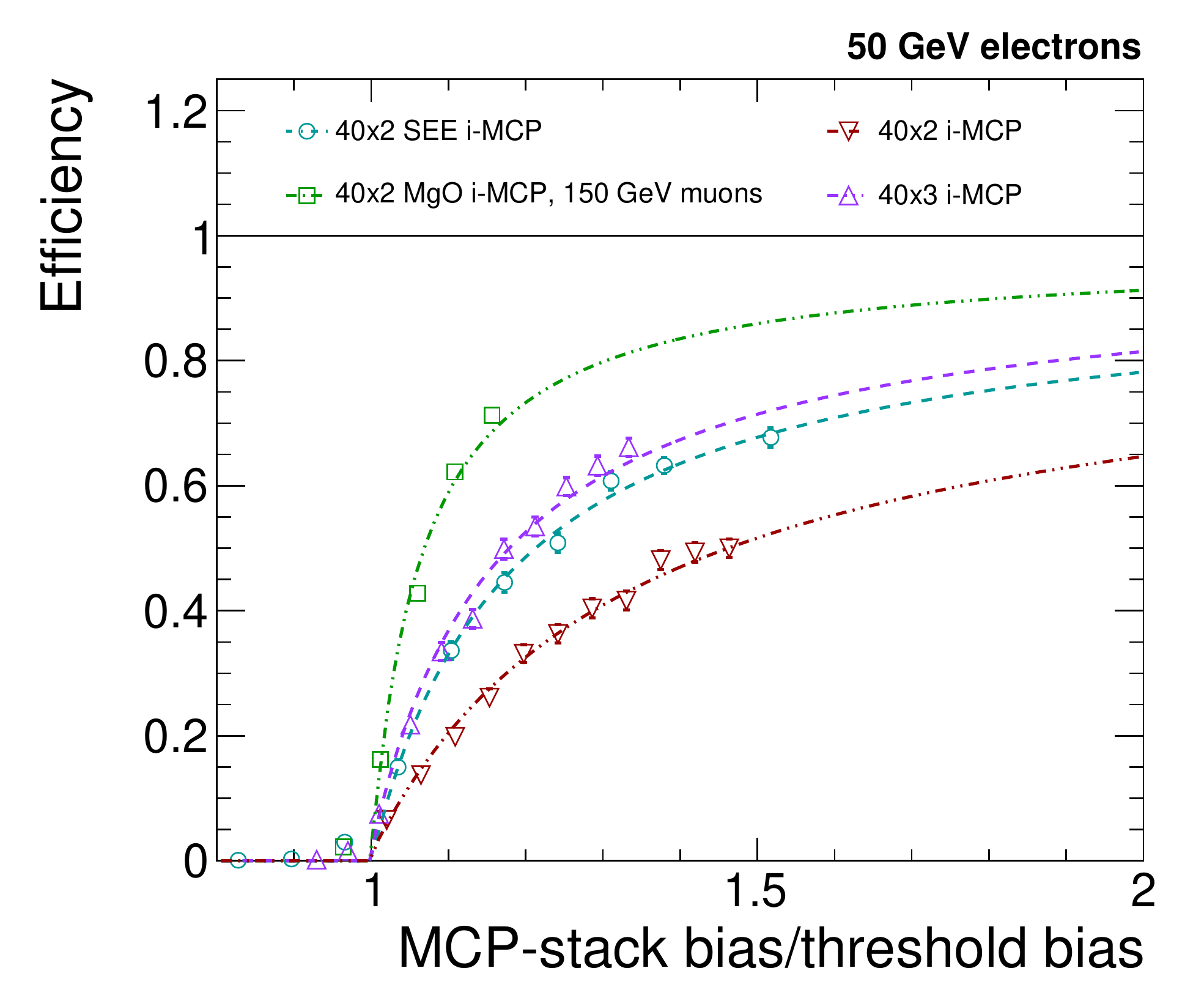}
\caption{Left) Efficiency to 50~GeV electrons and 150~GeV muons of MCP detectors with direct (full dots) and retarding bias (empty dots, squares and triangles)
between the photocathode and the first MCP layer as a function of the stack bias.
For the detectors operated in i-MCP mode the curve through the points is a fit to data of the response model detailed in the text (Eq.~\ref{model}).
For the detector operated in PMT-MCP mode data are fitted with a Fermi-Dirac function.
Right) Same efficiency as a function of the stack bias normalized to the threshold voltage, extracted from Eq.~\ref{model}.}
\label{fig:efficiencyH4}
\end{center}
\end{figure}

The efficiency of each MCP is defined as the fraction of events with
a signal above threshold with respect to the total number of single electron events, selected as discussed 
in the previous section.
The threshold is defined as 5 times the noise
of the detector, which is measured in pedestal runs or in events without electrons impacting on the scintillator.
The efficiency of some prototypes exposed to the H4 beam and operated in i-MCP mode is shown in Fig.~\ref{fig:efficiencyH4}
on the left as a function of the MCP-stack bias.
For a comparison, the efficiency of one of them operated in PMT-MCP mode is also shown.

In PMT-MCP mode the photoelectrons are extracted from the photocathode and amplified above
the detection threshold by the MCP. As it was found in~\cite{btf14}, a 100\% efficiency is obtained
at plateau, where the MCP gain is high enough to supply single photoelectron detection.
In i-MCP mode, the MCP layers have the dual function of initiating the cascade process and providing
signal amplification. Different channel lengths are involved in the amplification process, depending
on where the secondary electron is emitted, and inefficiencies may arise either because of
lack of secondary emission or because of insufficient amplification in the cascade following the secondary
emission. A model was developed in our earlier work~\cite{btf14} in which the detection efficiency
is parameterized as

\begin{equation}
\label{model}
\varepsilon = s \Big(1-\frac{1}{b \cdot \ln(V/V_{th})+1}\Big), ~~~~~~~~~~~~~~~~~~~~~~~~~(V \ge V_{th})
\end{equation}
under the hypothesis that the gain has a power-law dependence on the bias voltage.
The model parameters $V_{th}$, $s$ and $b$ are extracted from data.
The parameter $s$ represents the probability of secondary emission over the entire MCP length
and it is found to be compatible with one for the detectors under study.
The parameter $b$ describes the gain and depends on the device.
$V_{th}$ is the threshold voltage at which secondary electrons generated at the surface
of the MCP receive the exact gain needed for detection.
The behaviour of the efficiency versus the bias voltage can be explained with a change in the volume
of the sensitive region, that is the region where an electron receives enough amplification to be detected as a signal.
At the threshold voltage, the whole MCP length is needed in order to amplify the signal over a detectable threshold
and the signal is detected only if the secondary emission occurs at the surface of the MCP.
The efficiency increases with the bias voltage, since signals overcome the detection threshold also if the
secondary emission occurs inside the MCP layers.
The curves in Fig.~\ref{fig:efficiencyH4} are extrapolated to voltage values which can not be currently reached
due to the MCP electric strength.

The prototype with two layers (\emph{40$\times$2}) reaches a maximum efficiency of about 50\% for single electrons,
confirming the results in~\cite{btf14}.
Adding a third layer (\emph{40$\times$3}) the efficiency raises up to about 70\%,
since the  number of channel interfaces crossed by the particle is increased.
As suggested in~\cite{btf14}, and demonstrated in Fig.~\ref{fig:efficiencyH4}, the detection efficiency increases with the number of layers.
However, even with three layers the curve is not at plateau, suggesting that the efficiency gain is too slow.

The efficiency is larger, for the same channel length, in i-MCPs with enhanced secondary emission yield.
The \emph{SEE} and \emph{MGO} prototypes, also shown in Fig.~\ref{fig:efficiencyH4}, are two-layers i-MCPs.
The \emph{SEE} prototype underwent a surface treatment in order to increase the channel walls secondary 
emission efficiency.
The \emph{MGO} prototype is made of layers produced from a borosilicate glass substrate with an
atomic layer deposition technique (ALD by INCOM~Ltd~\cite{ald}\cite{ald2}) and a final layer of Magnesium oxide.
The \emph{SEE} device reaches an efficiency of about 50\% at a stack bias of about 1750~V, while for regular i-MCPs
with 2 layers the same efficiency is reached at about 3000~V. Also, the efficiency raises up to about 70\%.
The same efficiency is reached for the \emph{MGO} prototype. Owing to an accident while improving the test setup,
we could not study the response of the \emph{MGO} prototype at biases larger than 2400~V. The analysis however
shows that with MgO coatings efficiencies as high as 90\% could be attained still at moderate high voltages.

For all the i-MCP prototypes, data are well described by the model of Eq.~\ref{model}, suggesting that the 
efficiency trend with the voltage does not depend on the specific characteristics of the detector. 
The universality of the model for single stage i-MCPs
is further demonstrated in the right panel of Fig.~\ref{fig:efficiencyH4}, where the efficiency is shown 
as a function of the
stack bias normalized to the threshold voltage, extracted from a fit of the functional form of Eq.~\ref{model}
to the data.

\subsection{Efficiency of multi-stage MCP stacks}
\label{multistage}

An alternative way to improve the i-MCP behaviour consists in separating the region
where the signal is created from the amplification stage, providing a separate bias for the two layers.
This configuration is motivated by the fact that
a few photoelectrons entering two or three MCP layers give a clear signal above noise,
as demonstrated operating the detectors in PMT-MCP mode.

\begin{figure}[!htbp]
\begin{center}
\includegraphics[width=0.60\textwidth]{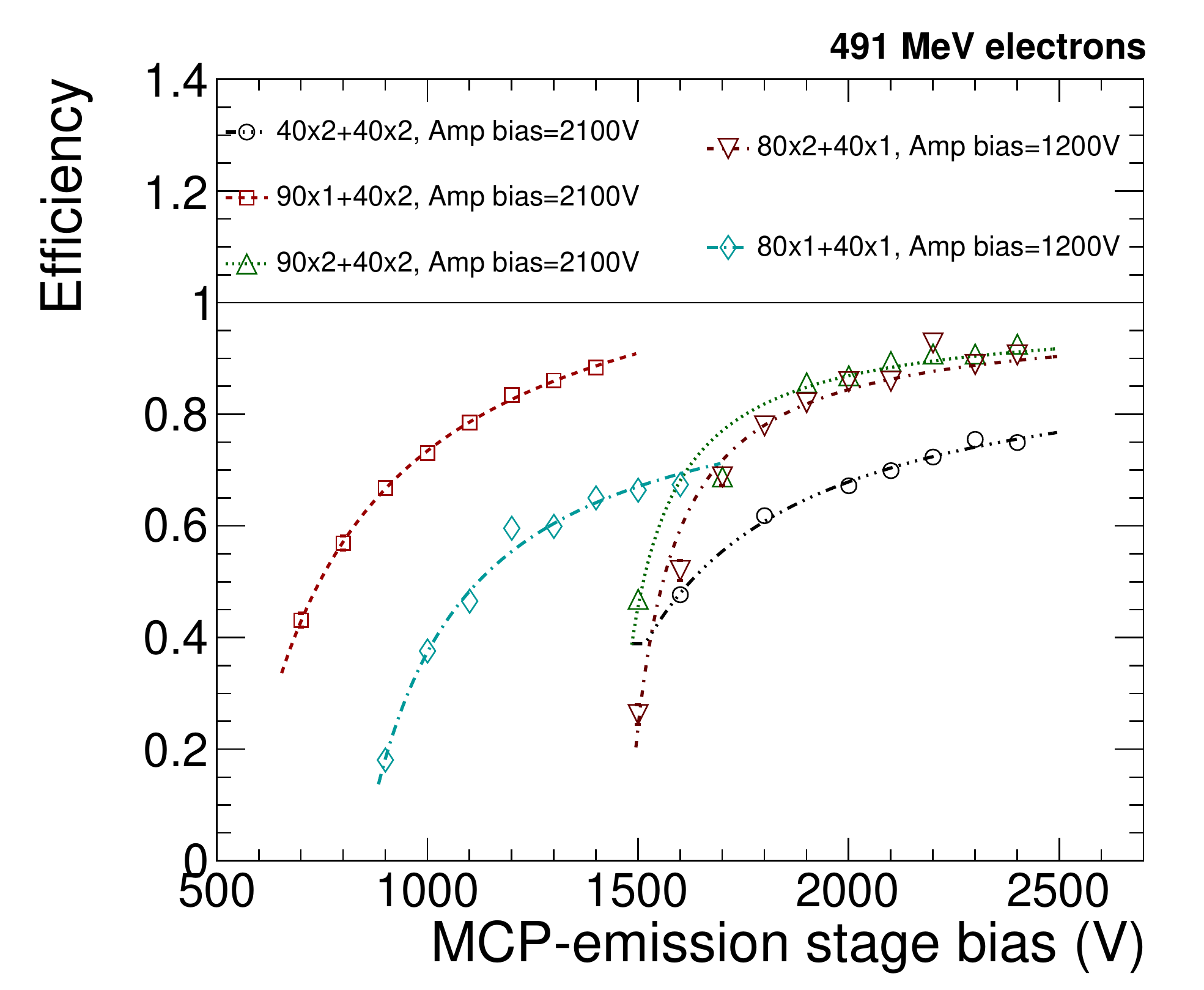}
\caption{Efficiency to 491~MeV electrons of multi-stage MCP detectors operated in i-MCP mode. The efficiency is given as a function of the emission stage bias, while the amplification stage\
 bias is reported in the legend. The curve through the points is a fit to data of the response model detailed in the text (Eq.~\ref{modelMS}).}
\label{fig:twostages}
\end{center}
\end{figure}

In Fig.~\ref{fig:twostages}
we show the efficiency of the multi-stage prototypes as a function of the MCP-stack bias, which is
varied to change the response of the devices. The amplification stage voltage is fixed.
For the detectors in the vacuum chamber (\emph{80$\times$1+40$\times$1}, \emph{80$\times$2+40$\times$1})
the probability for a single electron to originate a shower before reaching the MCP is not negligible.
A correction factor is therefore computed using the data collected at the T9 beam area, where one
of the i-MCPs was exposed to a mixed beam of electrons and pions.
The electron and pion components were separated with a Cherenkov detector
and the efficiency in response to each component was computed.
The i-MCP efficiency in response to electrons is 
larger than to pions, which means that also electrons produced in the primary electron cascade may generate a signal.
Describing the i-MCP response with a binomial per-event probability
\begin{equation}
\varepsilon_n = 1 - (1 - \varepsilon_1)^n,
\end{equation}
where $\varepsilon_1$ is the efficiency to single particles and $n$ is the multiplicity of electrons per electron
crossing the MCP, we find an average $n$ of about 1.4.
The efficiencies measured at the BTF on the i-MCPs inside vacuum chambers are corrected for this factor,
neglecting the difference in moving from 2~GeV to 491~MeV.

Three of the multi-stage MCPs have the same amplification region, consisting of two layers with
aspect ratio 40:1 (\emph{40$\times$2}) operated at a voltage around 2100~V.
The efficiency of these MCPs in response to charged particles does not reach zero even at low
stack bias, the residual efficiency being given by the amplification stage itself.
Experimental results suggest that the efficiency is
larger using layers with large aspect ratio in the emission stage
(\emph{90$\times$2+40$\times$2}, \emph{90$\times$1+40$\times$2}). 
An efficiency larger than 90\% is reached in the configuration \emph{90$\times$2+40$\times$2}, which is almost
20\% larger than with small aspect ratio, e.g. \emph{40$\times$2+40$\times$2}.
In the other two chambers the amplification region consists of one layer with aspect ratio 40:1 (\emph{40$\times$1}).
Also in this case, the efficiency increases by 20\% adding an extra layer in the emission stage
(\emph{80$\times$2+40$\times$1} with respect to \emph{80$\times$1+40$\times$1}).
In summary, the efficiency can be increased either building chambers with thicker layers
or adding extra layers. Since the prototypes \emph{40$\times$3}, discussed in Section~\ref{singlestage}, and \emph{80$\times$1+40$\times$1}
reach similar efficiencies, the relevant factor appears to be only the total thickness, independently of the number
of layers which are employed.
Finally, the prototypes \emph{90$\times$2+40$\times$2} and \emph{80$\times$2+40$\times$1}
reach the same plateau efficiency, while the efficiency
at low bias voltage is larger for the stack with an additional amplification layer.
This suggests that the addition of extra layers in the amplification stage gives only a small contribution to the total efficiency
if the efficiency of the emission stage is large enough.

The efficiency as a function of the emission stage bias (HV) for multi-stage i-MCPs is parameterized in
Fig.~\ref{fig:twostages} as
\begin{equation}
\label{modelMS}
\varepsilon = P_{amp}(1-P_{em}(HV))+P_{em}(HV)
\end{equation}
$P_{amp}$ is the contribution of the amplification layer only, which may have an efficiency larger than zero
even when no electron is generated in the emission layers and which does not depend on the voltage of the emission stage.
$P_{amp}$ is compatible with zero for the devices where the amplification stage is made of a single layer MCP,
and it is slightly above 20\% for the devices with double layer amplification stage.
$P_{em}$ may be parameterized as in Eq.~\ref{model}.
This model well describes the data as a function of the voltage of the emission stage, which suggests
that the parameterization in Eq.~\ref{model} works well for every single stage i-MCP detector.

\subsection{Time resolution}
\label{timeSingle}
The time resolution of i-MCP detectors exposed to a single particle beam is extracted comparing the time
of the detector under study with the time of a reference MCP operated in PMT-MCP mode and put upstream
of the i-MCPs along the beam line.
Pulses consistent with a single electron entering the test setup are selected in the same way as for
the efficiency measurement.
The signal time information is estimated with the CFD method 
and only the events where the time of the i-MCP detector is compatible with the time of the reference
PMT-MCP within 1~ns are retained for the analysis.

\begin{figure}[!htbp]
\begin{center}
\includegraphics[width=0.45\textwidth]{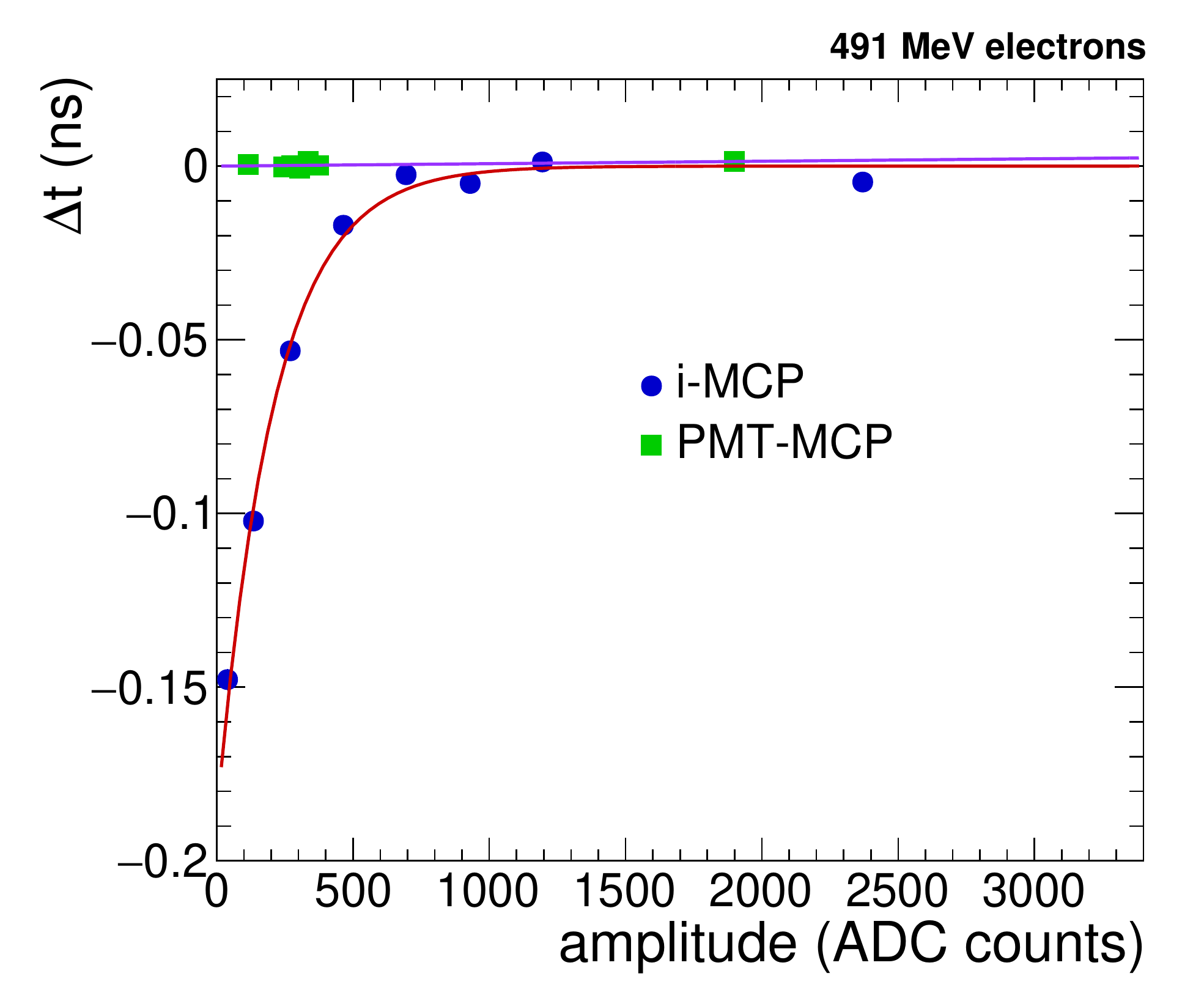}
\includegraphics[width=0.45\textwidth]{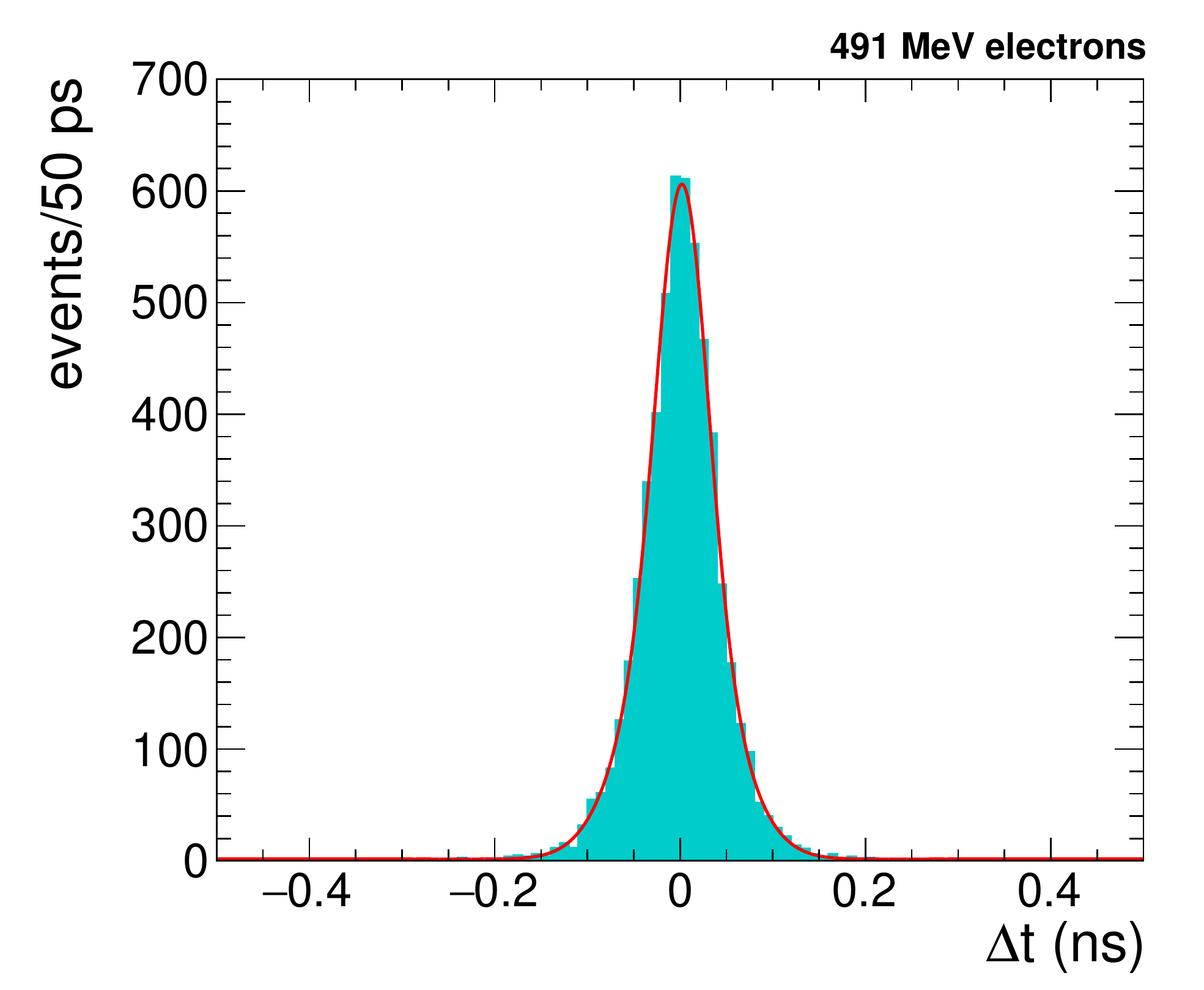}
\caption{Left: Difference between the time of a MCP detector, operated either in i-MCP mode (dots) or PMT-MCP mode (squares), and the time of a reference 
detector operated in PMT-MCP mode as a function of the signal amplitude. 
Right: Distribution of the time difference between the signals measured in two MCPs, one operated in i-MCP mode and the other one
in PMT-MCP mode. Both plots refer to MCPs exposed to 491~MeV electrons. }
\label{fig:corrNonLin}
\end{center}
\end{figure}

When a detector is operated in i-MCP mode, differently from the PMT-MCP mode case,
time non-linearities as a function of the signal amplitude arise.
The time difference between signals with low and large amplitude is compatible with a fraction of the transit time,
which is around 200-300~ps, therefore it is reasonable to assume that the signals with low amplitude are in average created
closer to the anode than larger signals.
This effect, which is shown in Fig.~\ref{fig:corrNonLin} on the left, is corrected for
with an empirical function.
The distribution of the time difference between the \emph{90$\times$1+40$\times$2} detector operated in i-MCP mode and the reference PMT-MCP,
after the non-linearity correction, is shown in Fig.~\ref{fig:corrNonLin} on the right in response to 491~MeV electrons and
for an operating voltage of 1400~V.
To compute the i-MCP time resolution a Gaussian fit to this distribution is performed, then 
the resolution of the reference PMT-MCP is subtracted. The latter is extracted comparing the relative resolution of the
time difference with respect to another PMT-MCP detector and with respect to an i-MCP detector. 
It is found to be 17$\pm$2~ps.

\begin{figure}[!htbp]
\begin{center}
\includegraphics[width=0.45\textwidth]{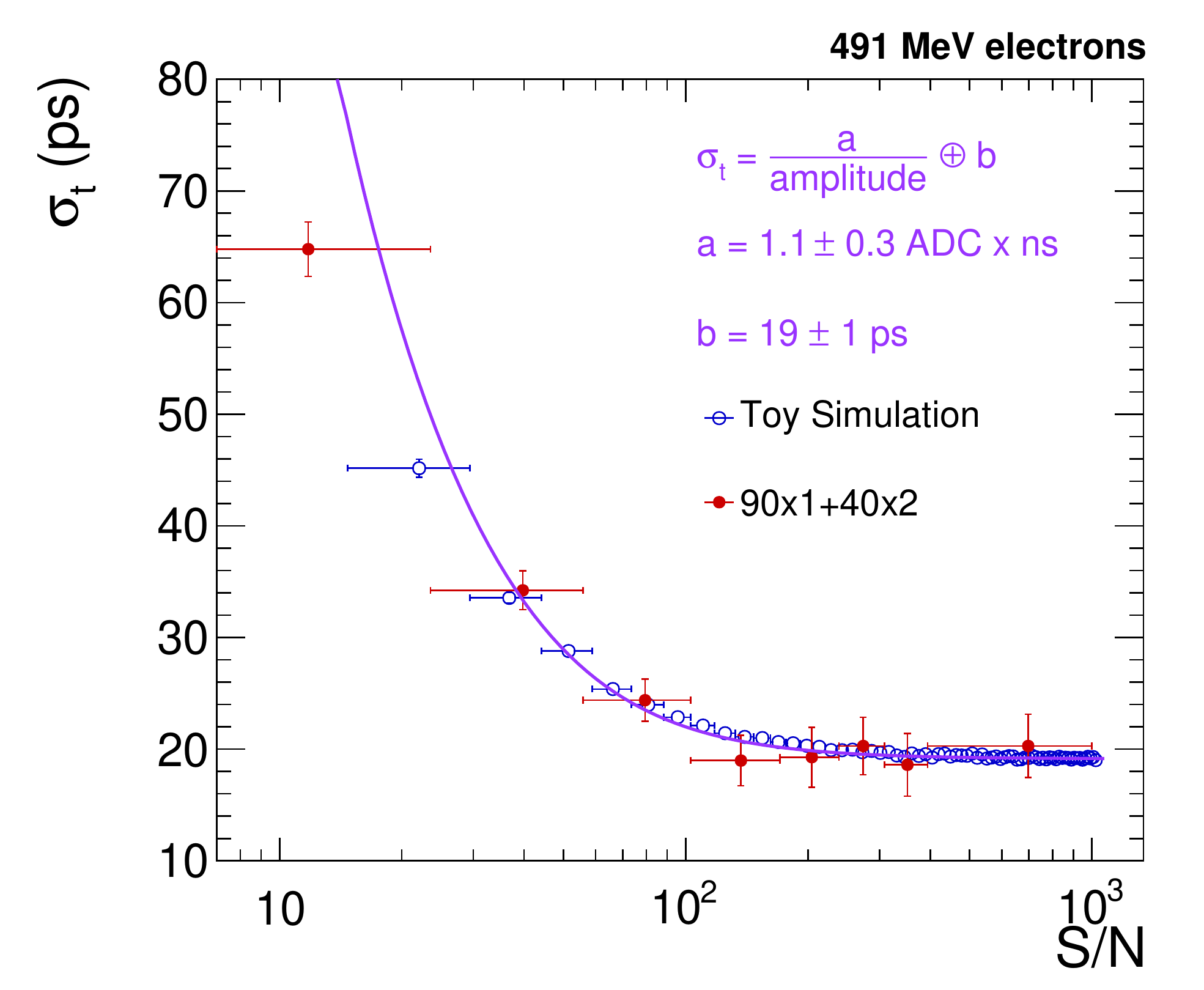}
\caption{Time resolution for an i-MCP prototype exposed to 491~MeV electrons as a function of the signal over noise ratio.
The resolution of the reference MCP operated in PMT-MCP mode is subtracted.}
\label{fig:timeResVsAmplitude}
\end{center}
\end{figure}

Fig.~\ref{fig:timeResVsAmplitude} shows the time resolution of the \emph{90$\times$1+40$\times$2} i-MCP detector
as a function of the signal over noise ratio, after the resolution of the reference PMT-MCP is subtracted.
The trend can be parametrized as
\begin{equation}
\sigma = \frac{a}{S/N} \oplus b,
\label{eqRes}
\end{equation}
from which the noise component $a$ and the constant term $b$ of the resolution are extracted.
The parameter $b$ determines the best resolution which can be achieved and it mainly depends on the transit time spread,
i.e. the time difference due to the different paths followed by the electrons in reaching the anode. Other sources can also
contribute, like imperfections in the time reconstruction or in the non-linearity corrections.
In Fig.~\ref{fig:timeResVsAmplitude} data are compared to a toy simulation
which is based on a pulse shape template built from data. Uncorrelated noise
is added to each sample as extracted from a pedestal run.
The noise contribution $a$ measured in data is compatible with the expectations from simulation.
The current simulation does not contain sources which can contribute to the constant term, so 
a Gaussian smearing is added to match the data.
The time resolution averaged over all the signal amplitudes is 26~$\pm$~2~ps for this device.
For a comparison, the time resolution for single photoelectron events of the same device operated in PMT-MCP 
mode and measured with a laser is 24~ps~\cite{barn}.

The average time resolution and the constant term of Eq.~\ref{eqRes} were computed for all the devices under test.
The results are compatible among the different prototypes and constant terms between 20~ps and 30~ps are obtained in all cases.
Only a small dependence of the time resolution on the number of layers and their aspect ratio is observed.
The data collected do not allow a systematic study of the time resolution dependence on the channel diameter
and layer thickness.

\section{Response to electromagnetic showers}
\label{showers}

\begin{figure}[!htbp]
\begin{center}
\includegraphics[width=0.49\textwidth]{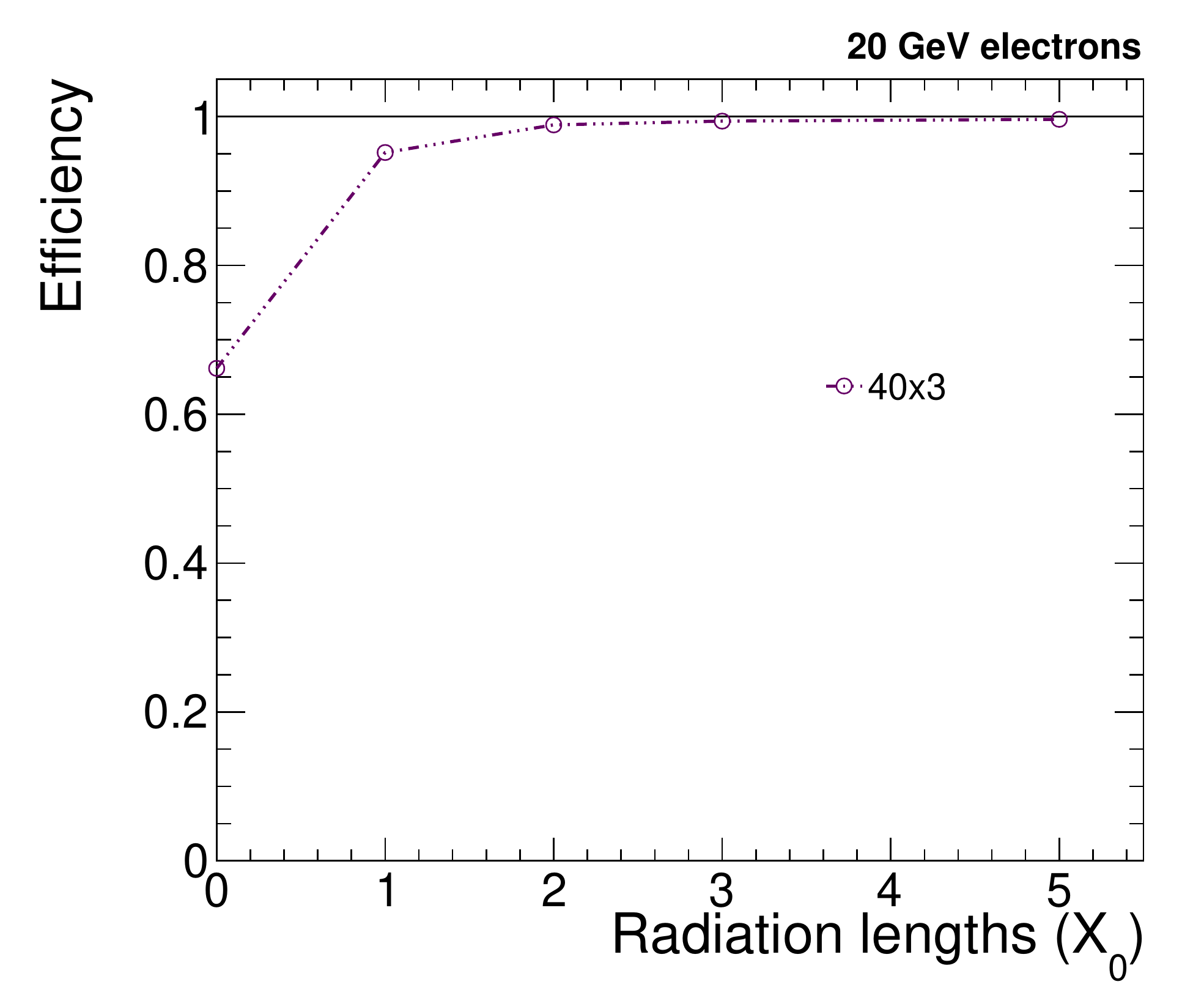}
\includegraphics[width=0.49\textwidth]{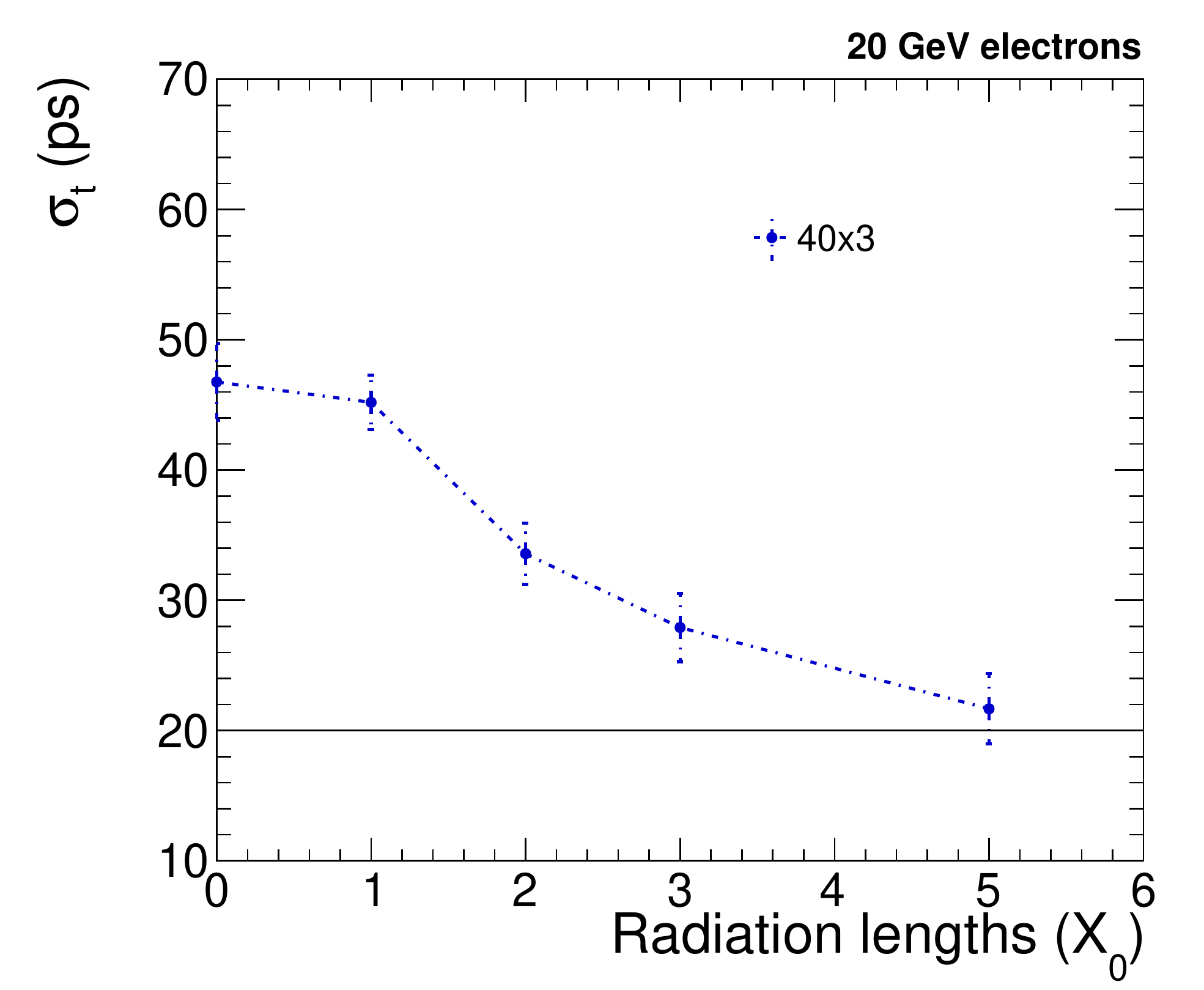}
\caption{Efficiency (left) and time resolution (right) of MCP prototypes operated in i-MCP mode and exposed to 20~GeV electrons
as a function of the absorber thickness, in units of radiation lengths. The contribution of the reference MCP to the time resolution is subtracted.}
\label{fig:showers}
\end{center}
\end{figure}

To characterize the i-MCP response to electromagnetic showers, data were acquired with a set of absorbers
of variable thickness and up to almost 5~radiation lengths $X_0$ in front of the detector.
The tests were performed at the CERN H4 Area.\\
The evolution of the efficiency to detect 20~GeV electrons is shown in Fig.~\ref{fig:showers} on the left for a 3 layers i-MCP
as a function of the absorber thickness in units of $X_0$. The efficiency definition
is the same as in Sec.~\ref{singleele}. The device was operated at the maximum voltage
used in the bias voltage scan performed in the single particle setup,
corresponding to the highest detection efficiency in that configuration.
The efficiency increases with the thickness of the absorber, reaching 95\% after 1$X_0$ and close to 100\% after 2$X_0$.
These results, which confirm the indications in~\cite{btf14}, are promising
in view of the usage of the i-MCP detectors in calorimeters at hadron colliders.

The time resolution in response to electromagnetic showers initiated by 20~GeV electrons
is shown as a function of the shower depth in Fig.~\ref{fig:showers} on the right, after the removal of the
contribution of the reference MCP resolution.
Due to the increased multiplicity of charged particles the digitizer input can saturate
and in this case the time to cross a fixed threshold of 500~ADC counts is used as time estimate, 
instead of the CFD algorithm.
The same event selection as in Sec.~\ref{timeSingle} is used.
Fig.~\ref{fig:showers} refers to a different detector with respect to the one used as an example in the previous section,
with a larger noise but comparable constant term. This explains why the time resolution at 0$X_0$
is worse than in Fig.~\ref{fig:timeResVsAmplitude} when averaging over all the signal amplitudes.
The time resolution improves as expected with the absorber thickness, because
the MCP layers are crossed by a larger multiplicity of particles.
An inclusive time resolution of about 20~ps is reached after 5$X_0$.

\section{Summary and outlook}
\label{summary}

We report on the response of microchannel plates to single relativistic particles and to electromagnetic showers. 
Several prototypes of MCPs used as secondary emission detectors 
were exposed to particle beams at the INFN-LNF Beam-Test Facility and at CERN.
Configurations with multiple MCP stacks, different geometries or layer coatings were compared, to investigate how 
these aspects affect the total amount
of secondary emission and the channel gain. In the tested configurations, detection efficiencies to single 
relativistic particles between 50\% and 90\% and constant terms for the time resolution between 20~ps and 30~ps are reached. 
Measurements with electromagnetic showers sampled at different depths show that, in presence of a large enough number of particles,
detection efficiencies up to 100\% can be reached with average time resolutions as good as 20~ps, independently of the geometry.
Present results suggest that this detection technique is suitable for a precise determination of the time of high energy photons and 
charged particles, and could help in the event reconstruction at high luminosity colliders.

\section{Acknowledgements}

We warmly thank R. Bertoni, R. Mazza, M. Nuccetelli and F. Pellegrino for the preparation of the experimental setup.
We are indebted to B. Buonomo, C. Di Giulio, L. Foggetta and P. Valente for their help with the setup of the beam facility at Frascati. 
We are grateful to the CERN PS and SPS accelerator teams for providing excellent beam quality.
This research program is carried out in the iMCP $R\&D$ project, funded by the Istituto Nazionale di Fisica Nucleare (INFN) in the 
Commissione Scientifica Nazionale 5 (CSN5). 
It has also received funding from the European Union Horizon 2020 research and innovation programme
under the Marie Sklodowska-Curie grant agreement No 707080.
The production of tested BINP design exemplars and caring out their measurements in Novosibirsk was supported by the Russian Science Foundation (Project no.16-12-10221).

\end{document}